\documentclass[twocolumn, pra]{revtex4}
\usepackage{amsmath}
\usepackage{amssymb}
\newcommand{\ket}[1]{|#1\rangle}

\newcommand{\ot}{\mathrm{otherwise}}
\newcommand{\mbf}[1]{\mathbf{#1}}

\begin{document}
\input epsf

\title{Entanglement swapping for generalized non-local correlations}

\author{Tony Short$^{a}$,
Sandu Popescu$^{a,b}$, Nicolas Gisin$^c$} \affiliation{ $^a$HH
Wills Physics Laboratory,
University of Bristol, Tyndall Avenue, Bristol, BS8 1TL, UK\\
$^b$Hewlett-Packard Laboratories, Stoke Gifford, Bristol BS12 6QZ,
UK\\ $^c$ Group of Applied Physics, University of Geneva, 20 rue
de l'Ecole-de-m\'{e}decine, CH-1211 Geneva 4, Switzerland.\\ }

\begin{abstract}
We consider an analogue of entanglement-swapping for a set of
black boxes with the most general non-local correlations
consistent with relativity (including correlations which are
stronger than any attainable in quantum theory). In an attempt to
incorporate this phenomenon, we consider expanding the space of
objects to include not only correlated boxes, but `couplers',
which are an analogue for boxes of measurements with entangled
eigenstates in quantum theory. Surprisingly, we find that no
couplers exist for two binary-input/binary-output boxes, and hence
that there is no analogue of entanglement-swapping for such boxes.
\end{abstract}

\maketitle

\section{Introduction}

One of the most surprising aspects of quantum theory is its
ability to yield \emph{non-local} correlations, which cannot be
explained by any local hidden-variable model \cite{bell, CHSH}.
These correlations do not allow superluminal signalling, but are
nevertheless useful in many information theoretic tasks \cite{qi1,
qi2}. An interesting question is whether quantum theory yields the
maximal amount of non-local correlations consistent with
causality. Perhaps surprisingly, it has been shown that this is
not the case \cite{popescu}, and that it is possible to construct
a theoretical system which does not allow superluminal signalling,
yet which is more non-local than quantum theory. Such a system can
achieve the maximal possible value of 4 for the
Clauser-Horne-Shimony-Holt (CHSH)\cite{CHSH} expression, compared
to $2\sqrt{2}$ for quantum theory (the Cirel'son bound
\cite{cirelson}), or 2 for any local hidden variable model.

Recently, the idea of non-local correlations stronger than those
attainable in quantum theory has received considerable interest:
Van Dam \cite{vandam} has shown that they allow any bipartite
communication complexity problem to be solved with only one bit of
communication. Wolf and Wullschleger \cite{wolf}, Buhrman et al.
\cite{bit-com}, and Short et al \cite{no-bit-com} have considered
whether they can be used for oblivious transfer and
bit-commitment. Cerf. et al. \cite{singlet-sim} have shown that
they can be used to efficiently simulate  measurements on a
quantum singlet state, and Barrett et al. \cite{barrett, barrett2}
and Jones and Masanes \cite{jones} have characterised and
considered the inter-convertibilty of different non-local
correlations.

To investigate the properties of general non-local no-signalling
correlations, we consider an abstract correlation-system composed
of a number of black boxes (subsystems) held by different parties,
each of which has an input (their measurement setting) and an
output (their measurement result)\footnote{Note that this setup
differs from that of some other authors (eg. \cite{barrett}), who
consider that all parties share a single extended black box.
Treating the subsystems as separate (yet correlated) boxes
provides us with a closer analogue to the quantum case, and allows
us to utilize the concept of measurement-induced `collapse' in
later sections. The mathematical formalism is the same in both
cases.}.  We represent the combined state of all of the boxes by
the conditional probability distribution for their outputs given
their inputs.

The non-local correlations achievable using correlated box-states
are analogous to those achievable using entangled quantum states.
It is therefore interesting to see what properties of entanglement
have analogues for these more general non-local correlations. A
property which does have such an analogue can be viewed as a
general property of non-local correlations (and therefore not
specifically quantum), while a property without such an analogue
is specific to quantum theory, and therefore may reveal why
quantum theory has the particular form  that it does.

In this paper, we consider the analogue of entanglement-swapping
for correlated box states, in which non-local correlations between
Alice and Bob, and between Bob and Charlie, are used to generate
non-local correlations between Alice and Charlie.

In an attempt to achieve this, we consider the possibility of
introducing a new class of objects called `couplers', which can
perform an analogue for boxes of measurements with entangled
eigenstates in quantum theory. We first consider a natural
potential coupler in detail, showing how it fails to provide a
consistent solution, then proceed to develop a general framework
with which to explore other possibilities. Surprisingly, we find
that no couplers exist for two binary-input/binary-output boxes,
and hence that there is no analogue of entanglement-swapping for
such boxes.

The structure of the paper is as follows: In section II we define
general correlated box states, and in section III we briefly
review quantum entanglement-swapping. In section IV, we attempt to
achieve an analogue of entanglement-swapping for non-locally
correlated boxes. However, we show this is impossible to achieve
using a sequence of conditional measurements on individual boxes.
In section V we introduce couplers, and consider both general and
specific cases, and in section VI we present our conclusions.

\section{Correlated no-signalling boxes} \label{box_sec}

\subsection{General case}

Consider a general multi-partite system composed of $N$ correlated
subsystems, each of which can be moved about freely. We represent
each subsystem by a black-box, which has an input (corresponding
to the choice of which measurement to perform on that subsystem),
and an output (which is the result of the chosen measurement). We
will assume that only one input can be made to each box, and that
the corresponding output is obtained immediately (without having
to wait for messages to travel between the boxes). Furthermore, we
assume that the probability of obtaining a given set of outputs
$\mathbf{O} = \{O_1, \ldots, O_N \}$ from a system of boxes
depends only on the inputs $\mathbf{I} = \{I_1, \ldots, I_N \}$
which are made to those boxes, and not on the timings of those
inputs (which would be reference-frame dependent). The
\emph{state} of the boxes can therefore be represented completely
by a conditional probability distribution $P(\mbf{O}|\mbf{I})$.

As the boxes can be moved to any point in space, and their inputs
applied at any time, the ability to transmit information using
them would allow superluminal signalling. We therefore require
that the boxes obey the following no-signalling condition: For all
partitions of the boxes into two disjoint sets held by a sender
(with inputs $\mbf{I_S}$ and outputs $\mbf{O_S}$) and a receiver
(with inputs $\mbf{I_R}$ and outputs $\mbf{O_R}$), there exists a
probability distribution $P( \mbf{O_R} | \mbf{I_R})$ such that
\begin{equation} \label{no-signalling-cond-eqn}
\sum_{\mbf{O_S}} P(\mbf{O}|\mbf{I}) = \sum_{\mbf{O_S}} P(\mbf{O_S}
\mbf{O_R} |\mbf{I_S} \mbf{I_R}) = P( \mbf{O_R} | \mbf{I_R} )
\end{equation}
for all $\mbf{O_R}, \mbf{I_R}, \mbf{I_S}$. This ensures that when
the sender and receiver are separated (and therefore the receiver
does not know $\mbf{O_S}$) the receiver can learn nothing about
the sender's inputs $\mbf{I_S}$. It is therefore impossible for
the sender to transmit information to the receiver using the
boxes.

The probability distribution $P( \mbf{O_R} | \mbf{I_R})$ plays an
analogous role to the reduced density matrix of the receiver's
subsystems in quantum theory\footnote{Note that when
$P(\mbf{O}|\mbf{I})$ satisfies the no-signalling condition, $P(
\mbf{O_R} | \mbf{I_R})$ will also satisfy it.  $P( \mbf{O_R} |
\mbf{I_R})$ is therefore an allowed correlated box state.},
providing the probabilities for his boxes' outputs when considered
independently.

\subsection{Two-box binary-input/binary-output states}

\begin{figure}
\epsfxsize=2.5truein \centerline{\epsffile{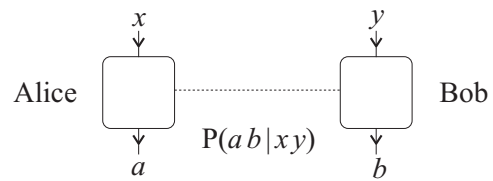}}
\caption{ Two correlated boxes held by Alice and Bob, in the state
$P(a b | x y)$. The dashed line between the two boxes represents
that they are non-locally correlated. \label{single-box-fig}}
\end{figure}

We now consider in detail the simplest case admitting non-local
correlations, which is that of two binary-input/binary-output
boxes. Taking $\mbf{I}=\{x, y\}$ and $\mbf{O}=\{a, b\}$, the two
box state is represented by the probability distribution $P(a b |
x y)$, where all of the inputs and outputs are binary variables.
We will assume that the first box is held by Alice and the second
box is held by Bob (as shown in figure \ref{single-box-fig}).

The no-signalling condition corresponds to the two requirements
\begin{eqnarray}
\sum_a P (a b | x y) &=&  P (b | y) \quad \forall x, y, b
\label{no-sig-rel1} \\
\sum_b P (a b | x y) &=&  P (a | x) \quad  \forall x, y, a
\label{no-sig-rel2}
\end{eqnarray}
for some pair of probability distributions $P (a | x)$ and $P(b |
y)$. The meaning of (\ref{no-sig-rel1}) is that Alice cannot
signal superluminally to Bob, while (\ref{no-sig-rel2}) means that
Bob cannot signal superluminally to Alice.

The complete class of probability distributions (and hence
2-box-states) consistent with relations (\ref{no-sig-rel1}) and
(\ref{no-sig-rel2}) has been investigated by Barrett et al.
\cite{barrett}, and was found to form an 8-dimensional convex
polytope with 24 vertices. Of these, 16 vertices represent the
deterministic `local' states, for which Alice and Bob's outputs
are a function of their inputs alone, and hence their boxes are
uncorrelated. These are the analogues of quantum mechanical
product states. The probability distributions for these extremal
local states are:
\begin{equation}
P^L_{\alpha \beta \gamma \delta} (a b | x y ) = \left\{
\begin{array}{ccl} 1 & :& a= \alpha x \oplus \beta \\ &&  b = \gamma y \oplus
\delta \\ 0 &:& \ot \end{array} \right.
\end{equation}
where $\alpha, \beta, \gamma, \delta \in \{0,1\}$ parameterize the
16 states and $\oplus$ denotes addition modulo 2. A state will lie
within the convex polytope $\mathcal{L}$ formed by these 16
vertices if and only if it can be simulated by local operations
and shared randomness, and we will refer to any such state as a
local state.

The remaining 8 vertices represent non-local states (lying outside
$\mathcal{L}$), with probability distributions given by:
\begin{equation} \label{nonlocal_box_eqn}
P^N_{\alpha\beta\gamma} (a b | x y) = \left\{
\begin{array}{ccl} 1/2 & :& a
\oplus b = x y \oplus \alpha x \oplus \beta y \oplus \gamma  \\
0&:& \ot
\end{array} \right.
\end{equation}
where $\alpha,\beta, \gamma \in \{0,1\}$. Each non-local extremal
state maximally violates a CHSH-type inequality (achieving a
greater value than is attainable in quantum theory), and cannot be
simulated with local operations and shared randomness. These
states are the analogues of maximally-entangled states in quantum
theory. Following \cite{barrett}, we will refer to all of the
non-local extremal states as PR-states \cite{popescu}, although
for simplicity we will usually consider the standard PR-state
$P^N_{000}(a b | x y)$, for which $a \oplus b = x y$.

\section{Entanglement-swapping in quantum theory}
\label{quantum-sec}

In the quantum case, the simplest example of entanglement-swapping
is as follows. Suppose that Alice shares a singlet state with Bob,
and Bob shares a singlet state with Charlie, such that their
combined state is
\begin{equation}
\ket{\Psi} = \frac{1}{2} \left( \ket{0}_A \ket{1}_{B_1} -
\ket{1}_A \ket{0}_{B_1} \right) \left( \ket{0}_{B_2} \ket{1}_C -
\ket{1}_{B_2} \ket{0}_C \right)
\end{equation}
In this state, Alice's and Charlie's qubits are completely
uncorrelated. However, expanding the bipartite states held by Bob,
and by Alice and Charlie, in the Bell basis of maximally entangled
states,
\begin{eqnarray}
\label{bell-basis1}
\ket{\psi^{\pm}} &=& \frac{1}{\sqrt{2}}(\ket{01}\pm\ket{10}) \\
\ket{\phi^{\pm}} &=& \frac{1}{\sqrt{2}}(\ket{00}\pm\ket{11})
\label{bell-basis2}
\end{eqnarray}
yields
\begin{equation}
\ket{\Psi} = \frac{1}{2} \left(\begin{array}{c}\;
\ket{\psi^+}_{B_1 B_2} \ket{\psi^+}_{AC} - \ket{\psi^-}_{B_1
B_2}\ket{\psi^-}_{AC} \\ - \ket{\phi^+}_{B_1 B_2}
\ket{\phi^+}_{AC} + \ket{\psi^-}_{B_1 B_2} \ket{\psi^-}_{AC}
\end{array}  \right)
\end{equation}
A measurement by Bob in the Bell-basis $\{ \ket{\psi^+}_{B_1
B_2}$, $ \ket{\psi^-}_{B_1 B_2}$, $\ket{\phi^+}_{B_1 B_2}$,
$\ket{\phi^-}_{B_1 B_2} \}$ will therefore leave Alice and Charlie
sharing the same maximally-entangled state as Bob, which contains
strong non-local correlations. Alice and Charlie will know which
entangled state they share as soon as Bob tells them his
measurement result. If they wish, they can transform this state
into a specific maximally entangled state (eg. $\ket{\psi^-}$) by
performing a local operation on one subsystem.

\section{Generalised non-locality swapping I}
\label{non-locality1}

We now consider the analogue of entanglement swapping for
generalised non-local correlations. Consider first the simplest
case in which Bob shares a standard PR-state $P^N_{000} (a b_1 | x
y_1)$ with Alice and an identical PR-state $P^N_{000} (b_2 c | y_2
z)$ with Charlie.

\begin{figure}
 \epsfxsize=3truein
\centerline{\epsffile{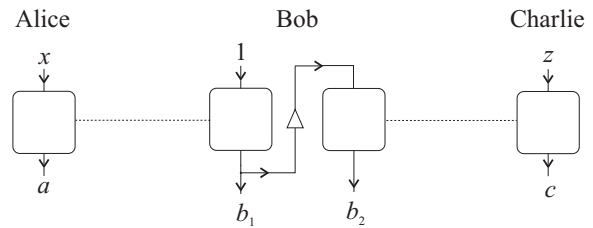}} \caption{Diagram showing
the two PR-states held by Alice and Bob, and by Bob and Charlie,
and a circuit applied by Bob to his two boxes. \label{wiring_fig}}
\end{figure}

The actions available to Bob are somewhat limited. The most
general thing that he can do is to apply an input to one of his
two boxes, and then use the output of this box in deciding which
input to apply to his second box. This can be represented by a
circuit diagram incorporating the two boxes (an example of which
is shown in figure \ref{wiring_fig}).

Can Bob introduce non-local correlations between Alice and
Charlie? Without loss of generality, suppose that Bob first inputs
$y_1=\lambda$ into his box which is correlated with Alice's box,
obtaining an output $b_1$. He can then input a general function
$y_2=\mu b_1 \oplus \nu$ of $b_1$ into his box which is correlated
with Charlie's box, obtaining the output $b_2$ (where $\lambda,
\mu, \nu, b_1, b_2 \in \{0,1\}$). An example circuit for
$\lambda=\mu=\nu=1$ is shown in figure \ref{wiring_fig}.
Regardless of the values of $\lambda, \mu$ and $\nu$, the
probability $P(\mathbf{b})$ of Bob obtaining the two-bit output
set $\mathbf{b}=\{b_1, b_2\}$ will be 1/4, as $P(b_1 | y_1) =
P(b_2 | y_2) =1/2$.

After completing this procedure, Bob announces his outputs $b_1$
and $b_2$ (and his strategy choices $\lambda, \mu$ and $\nu$) to
Alice and Charlie. The probability distribution for their inputs
and outputs is then given by
\begin{equation}
P'(a c | x z) = \left\{
\begin{array}{ccl} 1 & :& a= \lambda x \oplus b_1 \\ &&  c = (\mu b_1 \oplus \nu) z \oplus
b_2 \\ 0 &:& \ot \end{array} \right.
\end{equation}
which is the probability distribution corresponding to the
\emph{local} state $P^L_{\lambda b_1 (\mu b_1 \oplus \nu)  b_2} (a
c | x z)$. By applying inputs to his boxes, Bob has therefore
\emph{collapsed} the state of the remaining two boxes to an
extremal local state. This is analogous to the collapse of
entangled quantum states after a measurement by one party.

By switching the ordering of his inputs, selecting his strategy
choices ($\lambda, \mu, \nu$) probabilistically, restricting the
information he gives to Alice and Charlie, or sometimes announcing
that the process has `failed', Bob can cause the state shared by
Alice and Charlie to be any probabilistic combination of the
extremal local states (and hence any local state). However, there
is no way that Bob can introduce \emph{non-local} correlations
between Alice and Charlie, since in each particular instance of
the procedure Alice and Charlie share a local state.

The above result extends to the general case in which Bob shares
any correlated box state with Alice, and any correlated box state
with Charlie, but has no boxes which are correlated with both
parties. In this case, the initial state of all of the boxes will
be a product of two separate states:
\begin{equation}
P(\mbf{a} \mbf{b_1} \mbf{b_2} \mbf{c} | \mbf{x} \mbf{y_1}
\mbf{y_2} \mbf{z}) = P (\mbf{a} \mbf{b_1} | \mbf{x} \mbf{y_1}) P
(\mbf{b_2} \mbf{c} | \mbf{y_2} \mbf{z}),
\end{equation}
where $\mbf{x}$ and $\mbf{a}$, and $\mbf{z}$ and $\mbf{c}$,
represent the inputs and outputs of Alice's and Charlie's boxes
respectively, and Bob's boxes are partitioned into two sets (with
inputs $\mbf{y_1}$ and $\mbf{y_2}$, and outputs $\mbf{b_1}$ and
$\mbf{b_2}$ respectively) depending on whether they are correlated
with Alice or with Charlie.

The most general strategy Bob can adopt is to choose which of his
boxes to apply an input to, and what input to apply to that box,
dependent on all earlier outputs. Using such an approach, he will
generate a particular set of inputs and outputs with probability
$P(\mbf{b_1} \mbf{b_2} \mbf{y_1} \mbf{y_2})$. As Bob applies all
of his inputs first, $P(\mbf{b_1} \mbf{b_2} \mbf{y_1} \mbf{y_2})$
can be calculated without reference to Alice and Charlie, and will
depend only on the reduced state of Bob's boxes ($P(\mbf{b_1} |
\mbf{y_1})P(\mbf{b_2} | \mbf{y_2})$), and his particular choice of
strategy.

Unfortunately, once all of Bob's inputs and outputs are known, the
state $P(\mbf{a} \mbf{b_1} | \mbf{x} \mbf{y_1} )$ collapses to
$P(\mbf{a} |\mbf{x} \mbf{b_1} \mbf{y_1}) = P_{(\mbf{b_1},
\mbf{y_1})}(\mbf{a} |\mbf{x})$, which is a local probabilistic
operation for Alice, and similarly $P(\mbf{b_2} \mbf{c} |
\mbf{y_2} \mbf{z} )$ collapses to $P(\mbf{c} | \mbf{z} \mbf{b_2}
\mbf{y_2}) = P_{(\mbf{b_2}, \mbf{y_2})}(\mbf{c} |\mbf{z})$. For a
given set of inputs and outputs for Bob, the final state of
Alice's and Charlie's boxes will therefore take the form
\begin{equation}
P'(\mbf{a} \mbf{c} | \mbf{x} \mbf{z}) = P_{(\mbf{b_1},
\mbf{y_1})}( \mbf{a} | \mbf{x}) P_{(\mbf{b_2}, \mbf{y_2})}(
\mbf{c} | \mbf{z})
\end{equation}
which is manifestly local between Alice and Charlie. Regardless of
his strategy, it is therefore impossible for Bob to generate
non-local correlations between Alice and Charlie, even with some
small probability.

\section{Generalised non-locality swapping II: Couplers}

\subsection{A potential coupler}

\label{natural_coupler_sec}

As we have seen above, it seems impossible for Bob to generate
non-local correlations between Alice and Charlie. However we know
from quantum theory that such `entanglement-swapping' is possible.
Why then, can generalised non-locality not be swapped? Pondering
this question, one soon realises that the set of actions
considered above is far too restrictive. If we were to consider
analagous procedures in the case of quantum entanglement-swapping
(as introduced in section \ref{quantum-sec}), they would
correspond to Bob making an individual measurement on one of his
two qubits, and then performing a measurement on his second qubit
in a basis determined by his first measurement result. Whatever
Bob's results, such a procedure would collapse the state of
Alice's and Charlie's qubits into a local product state, and hence
cannot achieve entanglement-swapping.

\begin{figure}
\epsfxsize=3truein \centerline{\epsffile{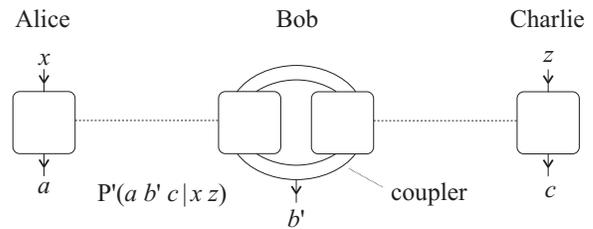}}
\caption{In order to generate non-local correlations between Alice
and Charlie, Bob attaches a coupler to the input and output
channels of his two boxes. The coupler generates a one-bit output
$b'$. \label{coupler_fig} }
\end{figure}

In quantum theory, entanglement swapping is achieved by a
\emph{joint} measurement on both of the subsystems held by Bob
(i.e. a measurement with entangled eigenstates). It is this
coupling of two subsystems which we need to incorporate into our
box model. We therefore need to introduce a new device called a
`coupler', which is connected to the input and output channels of
the two boxes held by Bob, and produces a single output $b'$ (as
shown in figure \ref{coupler_fig}). This output can be interpreted
as the result of a joint measurement on Bob's two subsystems.
\footnote{Note that we do not give the coupler an input, as we
intend it to correspond to a specific measurement (analogous to
the Bell measurement in quantum mechanics), whereas the standard
box inputs correspond to a selection of possible measurements.}.

Consider first the simple situation in which Alice and Bob share a
standard PR-state with probability distribution $P^N_{000}(a b_1 |
x y_1)$, and Bob and Charlie share a standard PR-state with
probability distribution $P^N_{000} (b_2 c | y_2 z)$. We would
like our coupler to generate, with some probability, a standard
PR-state $P^N_{000}(a c |x z)$ between Alice and Charlie.

As in the quantum case, such a process cannot be achieved with
certainty, as otherwise if Alice and Charlie brought their boxes
together Bob could signal superluminally to them by applying the
coupler (thereby changing Alice and Charlie's joint probability
distribution from $P(a c|x z)=1/4$ to $P'(a c | x z)=P^N_{000}(a c
|x z)$). However, if the coupler has binary output $b'$, we can
consider the case in which the probability distribution for two
coupled standard PR-states is
\begin{equation} \label{coupler_prob}
P'(a b' c | x z) =\left\{
\begin{array}{ccl} 1/4 & :& a
\oplus b' \oplus c = x z   \\
0&:& \ot
\end{array} \right.
\end{equation}
In this case, it is easy to see that each party's output is
random, and the outcomes of any two parties are uncorrelated (In
particular $P'(a c |x z)= P(a c |x z) = 1/4$ as required). Only by
learning all three outputs is any information about the inputs
obtained, and the coupler cannot therefore be used for signalling.

However, after Bob has applied the coupler and obtained an output
$b'$ (with probability $P'(b')=1/2)$ , Alice and Charlie will
share the maximally non-local state $P^N_{00b'}(a c |x z)$. If Bob
then announces his measurement result, and one party performs the
local operation $a \rightarrow a \oplus b'$ on their output (i.e.
applies a NOT-gate if $b'=1$), Alice and Charlie will be left with
the standard PR-state $P^N_{000}(a c |x z)$ as desired. This
procedure, in which Alice or Charlie must perform a local
correction conditional on Bob's measurement result in order to
obtain a desired final state, is strongly analogous to the quantum
case.

It therefore seems that, by enlarging the class of generalised
non-local objects to include couplers in addition to boxes, we
achieve generalised non-locality swapping. This is in complete
analogy with quantum mechanics where in addition to entangled
states, we consider measurements with entangled eigenstates.
However, as we will show below, the above coupler actually cannot
exist.

\subsection{Difficulties with the potential coupler}

Before allowing the coupler defined in the last section in our
model, it is important to check that it gives consistent results
when applied to all possible states. To investigate this, we
consider the case in which Alice and Charlie apply inputs to their
boxes before Bob applies the coupler to his boxes. If they
initially share standard PR-states as before, their inputs and
outputs will obey the relations $a \oplus b_1 = x y_1$ and $b_2
\oplus c = y_2 z$. After Alice and Charlie have applied inputs to
and obtained outputs from their boxes, the outputs of Bob's two
boxes will be given by $b_1 = x y_1 \oplus a$ and $b_2 = z y_2
\oplus c$. The probability distribution for Bob's two boxes has
therefore collapsed to the extremal local state $P^L_{xazc}(b_1
b_2 | y_1 y_2)$.

Suppose that Bob then applies a coupler to his two boxes. As is
the case for the original box inputs, we assume that the final
probability distribution given by (\ref{coupler_prob}) will be the
same regardless of the timings of Alice and Charlie's inputs, and
of Bob's application of the coupler. In order to satisfy
(\ref{coupler_prob}), it is therefore necessary that the
probability distribution $P'(b')$ for the coupler output when it
is applied to two boxes in the state
$P^L_{\alpha\beta\gamma\delta}(b_1 b_2 | y_1 y_2)$ be given by
\begin{equation} \label{local_box_eqn}
P'(b') = \left\{
\begin{array}{ccl} 1 & :& b' = \alpha \gamma \oplus \beta \oplus \delta \\
0&:& \ot
\end{array} \right.
\end{equation}

Similarly, in the case in which Alice and Charlie measure first,
all possible pairs of initial bipartite extremal-states shared by
Alice-Bob and Bob-Charlie (e.g. $P^N_{001}(a b_1 | x y_1)$ and
$P^L_{0110} (b_2 c | y_2 z)$) will collapse to a local extremal
state for Bob's two boxes ( e.g. $P^L_{x(a \oplus 1)01}(b_1 b_2 |
y_1 y_2)$ for the example given). If we assume that the coupler
always acts in the same way when applied to the same state of
Bob's two boxes, then equation (\ref{local_box_eqn}) can be used
deduce the action of the coupler on all initial states of this
type.

Furthermore, if we assume linearity in the initial probability
distributions (which is necessary to ensure no-signalling - as
shown in section \ref{dynamics_sec}), then we can deduce the
action of the coupler when applied between any two general states
by expanding them as convex mixtures of extremal states and using
the above results. It appears that such a coupler is consistent.

\begin{figure}
\epsfxsize=3.4truein \centerline{\epsffile{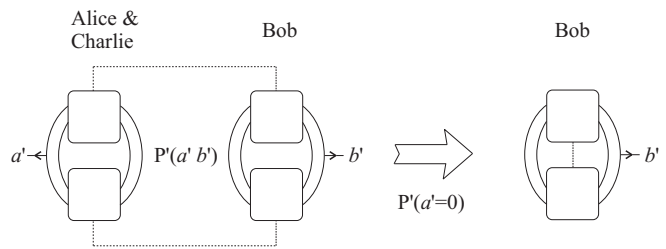}}
\caption{Alice and Charlie bring their boxes together and apply a
coupler between then, whilst Bob applies his coupler as normal. If
Alice and Charlie applies their coupler first and obtain $a'=0$,
Bob's two boxes will collapse into a standard PR-state. In this
case, Bob's coupler will output the same result as if it had been
applied directly to a PR-state \label{looped_coupler_fig}.}
\end{figure}

However, suppose that in addition to Bob applying a coupler to his
two boxes, Alice and Charlie bring their two boxes together and
apply a coupler between them (as shown in figure
\ref{looped_coupler_fig}). If Alice and Charlie's coupler is
applied first and they obtain output $a'$, Bob's two boxes will
collapse to the PR-state $P^N_{00a'}(b_1 b_2 |y_1 y_2)$ (from
equation (\ref{coupler_prob})). In order to obtain Bob's output,
we must therefore determine the coupler output when it is applied
to two boxes in an PR-state, rather than to two boxes in a local
state as we have considered so far.

To investigate this case, we consider a particular non-extremal
state $P^B (b_1 b_2 | y_1 y_2)$. As for mixed states in quantum
theory, every non-extremal box state can be obtained in an
infinite number of different ways by taking probabilistic mixtures
of other box states \cite{barrett}. We require that the coupler
act in the same way regardless of how the probability distribution
$P^B (b_1 b_2 | y_1 y_2)$ was prepared (i.e. it should be
decomposition invariant). The state $P^B (b_1 b_2 | y_1 y_2)$ has
two particular decompositions of interest. The first is obtained
by adding random noise to the standard PR-state until it becomes
local, and the second of which is its explicit decomposition in
terms of local extremal states:
\begin{eqnarray}
P^B(b_1 b_2 | y_1 y_2)  &=&  \frac{1}{2}\, P^N_{000}(b_1 b_2 | y_1
y_2)  \nonumber \\ &  &+ \frac{1}{8} \sum_{\beta\delta} P^L_{0
\beta 0 \delta} (b_1 b_2 | y_1 y_2)  \label{box_decomp1}
\\ &=& \frac{1}{8}
\sum_{\alpha \beta \gamma} P^L_{\alpha\beta\gamma(\alpha \gamma
\oplus \beta)}(b_1 b_2 | y_1 y_2) \label{box_decomp2}
\end{eqnarray}

Using (\ref{local_box_eqn}), it is easy to see that the eight
local states in the decomposition (\ref{box_decomp2}) of $P^B (b_1
b_2 | y_1 y_2)$ all give $b'=0$ when the coupler is applied to
them. Applying linearity, the coupler must therefore give $b'=0$
with certainty when applied to the state $P^B (b_1 b_2 | y_1
y_2)$.

However, it is also evident from (\ref{local_box_eqn}) that the
coupler will give $b'=1$ when applied to the local states
$P^L_{0001} (b_1 b_2 | y_1 y_2)$ and $P^L_{0100}(b_1 b_2 | y_1
y_2)$, which appear in the decomposition (\ref{box_decomp1}) with
probability 1/8 each. Given any non-negative probability
distribution ${P'}^N(b')$ for the coupler output when applied to a
standard PR-state, the probability of obtaining $b'=1$ from the
state $P^B (b_1 b_2 | y_1 y_2)$ cannot therefore be less than
$1/4$.

The results obtained from the two decompositions of $P^B (b_1 b_2
| y_1 y_2)$ are therefore inconsistent, and cannot be reconciled
by any physical probability distribution ${P'}^N(b')$. Only the
non-physical distribution
\begin{equation}
{P'}^N(b') = \frac{3}{2} - 2b',
\end{equation}
which is not a valid probability distribution as it does not
satisfy $0 \leq {P'}^N(b')\leq 1$, could recover the desired
decomposition invariance.

Because of this inconsistency, the `naive' coupler defined by
(\ref{coupler_prob}) is not an allowable object within the
correlated-box model. However, this raises the question of whether
other any  coupler exists which can achieve an analogue of
entanglement swapping.

\subsection{General couplers} \label{dynamics_sec}

To generalise the approach of the previous section, we will
consider a coupler as any device which acts on $n$ boxes with a
given range of inputs and outputs and produces a single output
$b'$ (also with a given range), and which cannot be implemented by
applying a sequence of individual inputs to the coupled boxes (as
in section  \ref{non-locality1}). In this context, the coupler
considered in the last section, were it to have proved consistent,
would have been an $n=2$ coupler, with all inputs and outputs
being binary.

We consider a general set of $N$ boxes divided between the person
who is going to apply the coupler (who we call Bob) and the rest
of the world (which we call Alice). In the most general case, Bob
has the $n$ boxes to which the coupler is to be applied (with
inputs $\mbf{y}$ and outputs $\mbf{b}$), and an additional set of
$m$ boxes (with inputs $\mbf{\check{y}}$ and outputs
$\mbf{\check{b}}$). Alice has the remaining $(N-n-m)$ boxes (with
inputs $\mbf{x}$ and outputs $\mbf{a}$). In the last section, for
example: $\mbf{x}= \{x, z\}$, $\mbf{a}= \{a, c\}$, $\mbf{y}=\{y_1,
y_2\}$, $\mbf{b}= \{b_1, b_2\}$,
$\mbf{\check{b}}=\mbf{\check{y}}=\{\}$.

The coupler then performs the transformation:
\begin{equation}
P(\mbf{a} \mbf{\check{b}} \mbf{b} | \mbf{x} \mbf{\check{y}}
\mbf{y}) \rightarrow P'(\mbf{a} \mbf{\check{b}} b' | \mbf{x}
\mbf{\check{y}})
\end{equation}

Following the discussion in the previous section, we impose four
natural constraints on the coupler's action:

\renewcommand{\labelenumi}{\roman{enumi}}

\begin{enumerate}
\item{\textbf{Universality}}: The coupler must be
applicable to any set of $n$ boxes with the appropriate range of
inputs/outputs, that are part of any no-signalling correlated box
state. Note that this is the condition for which the coupler
proposed in the last section fails, as it cannot be applied to two
boxes in a PR-state. \label{cond_u}

\item{\textbf{Completeness}}: A correlated box state is completely
specified by the conditional probability distribution for its
outputs given its inputs. To respect this completeness, we require
that the  probability distribution $P'(\mbf{a}\mbf{\check{b}} b' |
\mbf{x}\mbf{\check{y}})$ obtained by applying the coupler to a set
of boxes depends only on the probability distribution $P(\mbf{a}
\mbf{\check{b}} \mbf{b} | \mbf{x} \mbf{\check{y}} \mbf{y})$ of
those boxes to which it is applied.

The same probability distribution $P(\mbf{a} \mbf{\check{b}}
\mbf{b} | \mbf{x} \mbf{\check{y}} \mbf{y})$ can be obtained from
many different mixtures of extremal states (as in
(\ref{box_decomp1}) and (\ref{box_decomp2})), or by collapsing a
larger state by applying inputs to some of the boxes. This
requirement ensures that the coupler gives the same outcome in all
of these cases. It also ensures that the results do not depend on
the time at which the coupler is applied, just as the timings of
standard inputs do not affect the boxes' outputs.\label{cond_nhv}

\item{\textbf{No signalling}}: \label{cond_ns}
The coupler must not allow signalling. Note that the most powerful
situations for sending and receiving information are when all of
the boxes that are not held by Bob (i.e. Alice's boxes) are
gathered in the same place, and hence all of their inputs and
outputs are immediately accessible. We must rule out two
possibilities:
\begin{enumerate}
\item{Signalling from Alice to Bob}: We require that Bob
cannot learn anything about Alice's inputs from his coupler and
box outputs. We therefore require that: \label{cond_ns2}
\begin{equation} \label{no_coupler_sig_eqn2}
\sum_{\mbf{a}} P'(\mbf{a} \mbf{\check{b}} b' | \mbf{x}
\mbf{\check{y}} ) = P'( \mbf{\check{b}} b' | \mbf{\check{y}} )
\end{equation}
for some probability distribution  $P'( \mbf{\check{b}} b' |
\mbf{\check{y}} )$ which is independent of $\mbf{x}$.

\item{Signalling from Bob to Alice}: We require that Alice cannot
learn anything about Bob's box inputs, or about whether he has (or
has not) applied his coupler. We therefore require that:
\label{cond_ns1}
\begin{eqnarray}
&& \sum_{\mbf{\check{b}}, b' } P'(\mbf{a} \mbf{\check{b}}  b' |
\mbf{x} \mbf{\check{y}} ) = \sum_{\mbf{\check{b}}, \mbf{b} }
P(\mbf{a} \mbf{\check{b}}  \mbf{b} | \mbf{x} \mbf{\check{y}}
\mbf{y}) = P(\mbf{a} | \mbf{x}). \label{no_coupler_sig_eqn1}
\end{eqnarray}

\end{enumerate}

\item{\textbf{Non-triviality}}: The coupler must represent
something that was not previously possible. As discussed in
section \ref{non-locality1}, the most general strategy that Bob
can adopt \emph{without} a coupler is to apply a sequence of
individual inputs to his boxes, where later inputs may depend on
earlier outputs. A coupler cannot be simulated using such a
procedure (with $b'$ given by some function of the outputs).

\end{enumerate}

We will now show that constraints (i)-(iii) allow only those
couplers which act as linear maps on the reduced state of Bob's
boxes.

Let us consider the case in which Alice applies her inputs
$\mbf{x}$, and Bob applies his inputs $\mbf{\check{y}}$
\emph{before} he applies the coupler. They will obtain outputs
$\mbf{a}$ and $\mbf{\check{b}}$ respectively with probability
$P(\mbf{a} \mbf{\check{b}}| \mbf{x} \mbf{\check{y}})$, and the
state of Bob's boxes will collapse to a specific state
$P_{(\mbf{x}\mbf{a}\mbf{\check{y}}\mbf{\check{b}})}
(\mbf{b}|\mbf{y}) = P(\mbf{b} | \mbf{y} \mbf{x} \mbf{a}
\mbf{\check{y}} \mbf{\check{b}})$, which depends on the inputs and
outputs obtained.

From the completeness constraint (ii), the coupler will then act
on Bob's $n$-box reduced state exactly as it would have if that
$n$-box state were prepared directly (without collapsing a larger
$N$ box system), giving the output probability distribution:
\begin{equation}\label{two_term_coup_eqn}
P'_{(\mbf{x}\mbf{a}\mbf{\check{y}}\mbf{\check{b}})} (b') =
\mathcal{C} \left[ P_{(\mbf{x} \mbf{a} \mbf{\check{y}}
\mbf{\check{b}})}(\mbf{b} | \mbf{y} ) \right]
\end{equation}
where $\mathcal{C}$ is some function characteristic of the
coupler. The final probability distribution for Alice's and Bob's
outputs must therefore given by
\begin{equation} \label{general_big_box_eqn}
P'(\mbf{a}\mbf{\check{b}} b' | \mbf{x}\mbf{\check{y}}) =
P(\mbf{a}\mbf{\check{b}}| \mbf{x}\mbf{\check{y}})
P'_{(\mbf{x}\mbf{a}\mbf{\check{y}}\mbf{\check{b}})} (b').
\end{equation}

Note that as long as $P'_{(\mbf{x} \mbf{a} \mbf{\check{y}}
\mbf{\check{b}})}(b' )$ is a valid probability distribution
(satisfying $\sum_{b'} P'_{(\mbf{x} \mbf{a} \mbf{\check{y}}
\mbf{\check{b}})}(b')=1 $), equation (\ref{general_big_box_eqn})
will always satisfy the no-signalling constraint (iiib).

To investigate the class of allowed coupler functions
$\mathcal{C}$, and to incorporate the no-signalling constraint
(iiia), it is helpful to consider a particular class of
$(n+1)$-box states, for which Alice has a single box with input
$x$ and output $a$ and Bob has only those $n$ boxes to which he
will apply the coupler. The states we will consider are given by
\begin{equation} \label{3-term_box_eqn}
P(a \mbf{b}  | x  \mbf{y}) = \left\{
\begin{array}{ccl}  \lambda_a P_a (\mbf{b} | \mbf{y})  & : & x=0 \\
\sum_{a'} \lambda_{a'} P_a' (\mbf{b} | \mbf{y}) &:& x=1, a=0 \\  0
& : & \ot
\end{array} \right.
\end{equation}
where $P_a (\mbf{b} | \mbf{y})$ are a set of no-signalling $n$-box
states labelled by $a$, and $\lambda_a$ is the probability for
Alice to obtain output $a$ when $x=0$ (satisfying $\lambda_a >0$
and $\sum_a \lambda_a =1$). It is easy to check that $P(a \mbf{b}
| x \mbf{y})$ is a valid no-signalling state. A state of this type
corresponds to the case in which Bob is given an $n$-box state
selected randomly from some set, and Alice can either discover
which state Bob has been given (by inputting $x=0$ into her box)
or not (by inputting $x=1$). In the latter case, the collapsed
state of Bob's boxes will be a probabilistic mixture of all of the
boxes in the set.

Applying a coupler to Bob's boxes (which must be possible due to
the universality constraint (i)) yields the state:
\begin{equation}
P'(a b'| x) = \left\{
\begin{array}{ccl}  \lambda_a \mathcal{C} [P_a (\mbf{b} | \mbf{y})]  & : & x=0 \\
\mathcal{C} [\sum_{a'} \lambda_{a'} P_{a'} (\mbf{b} | \mbf{y})] &:& x=1, a=0 \\
0 & : & \ot
\end{array} \right.
\end{equation}
which is no-signalling from Alice to Bob, as required by
constraint (iiia), only when
\begin{equation}
\sum_a \lambda_a \mathcal{C} [P_a (\mbf{b} | \mbf{y})] =
\mathcal{C} [\sum_{a'} \lambda_{a'} P_{a'} (\mbf{b} | \mbf{y})] =
P'(b').
\end{equation}
In order for this relation to be satisfied for all choices of
$\lambda_a$ and $P_a (\mbf{b} | \mbf{y})$, $\mathcal{C}$ must be a
\emph{linear} function of Bob's $n$-box probability distribution,
of which the most general form is given by
\begin{equation}
P'(b') =  \sum_{\mbf{b} \mbf{y}} \chi(b', \mbf{b} \mbf{y}) P
(\mbf{b} | \mbf{y})  + \xi( b'). \label{complex_eqn}
\end{equation}
As $\sum_{\mbf{b}} P (\mbf{b} | \mbf{y})=1$, we can always
eliminate $\xi( b')$ by adding it to each of the coefficients
$\chi (b', \mbf{b} \mbf{y_0})$ for a particular $\mbf{y_0}$, hence
(\ref{complex_eqn}) can be simplified to give the final coupler
function
\begin{equation} \label{linear_coup_eqn}
P'(b')= \sum_{\mbf{b} \mbf{y}} \chi(b', \mbf{b} \mbf{y}) P
(\mbf{b} | \mbf{y}).
\end{equation}

Combining (\ref{general_big_box_eqn}) and (\ref{linear_coup_eqn}),
and using the fact that
\begin{equation}
P(\mbf{a} \mbf{\check{b}} \mbf{b} | \mbf{x} \mbf{\check{y}}
\mbf{y}) = P(\mbf{a} \mbf{\check{b}}| \mbf{x}
\mbf{\check{y}})P_{(\mbf{a}\mbf{x}
\mbf{\check{y}}\mbf{\check{b}})}  ( \mbf{b} | \mbf{y}),
\end{equation}
we find that the effect of the coupler on a general
box state is given by
\begin{equation}
P'(\mbf{a} \mbf{\check{b}} b' | \mbf{x} \mbf{\check{y}})
=\sum_{\mbf{b} \mbf{y}} \chi(b', \mbf{b} \mbf{y}) P(\mbf{a}
\mbf{\check{b}} \mbf{b}| \mbf{x}\mbf{\check{y}} \mbf{y}).
\label{general_big_box_eqn2}
\end{equation}

Although we have so far only considered the no-signalling
constraint (iiia) for a specific set of initial boxes given by
(\ref{3-term_box_eqn}), it is easy to see that
(\ref{general_big_box_eqn2}) will obey (iiia) for \emph{any}
initial state. As the initial distribution $P(\mbf{a}
\mbf{\check{b}} \mbf{b} | \mbf{x} \mbf{\check{y}}\mbf{y})$ is
no-signalling we have
\begin{eqnarray}
\sum_{\mbf{a}} P'(\mbf{a} \mbf{\check{b}} b' | \mbf{x}
\mbf{\check{y}} ) &=& \sum_{\mbf{b} \mbf{y}} \chi(b', \mbf{b}
\mbf{y}) P(\mbf{\check{b}} \mbf{b} | \mbf{\check{y}} \mbf{y})
\\ &=& P'(\mbf{\check{b}} b' | \mbf{\check{y}})
\end{eqnarray}
as required by (\ref{no_coupler_sig_eqn2}). Hence any coupler
obeying (\ref{linear_coup_eqn}) cannot be used for signalling.

The only remaining constraints on $\chi(b', \mbf{b} \mbf{y})$ are
universality (i) and non-triviality (iv). The former requires that
$P'(b')$ must be a valid probability distribution (satisfying
$P'(b')>0$ and $\sum_{b'} P'(b')=1$) for all initial $n$-box
states $P (\mbf{b} | \mbf{y})$ to which the coupler can be
applied, and the latter requires that the action of the coupler
cannot be simulated using Bob's standard box inputs.

\subsection{Two box binary-input/binary-output couplers }
\label{two-box-coupler-sec}

As a specific case of the general couplers introduced in the last
section, we will investigate the class of couplers which act on
two binary-input/binary-output boxes and generate a binary output
$b'$. Following equation (\ref{linear_coup_eqn}), if a coupler
characterised by $\chi(b', b_1 b_2 y_1 y_2)$ is applied to the
two-box state $P(b_1 b_2 | y_1 y_2)$, the final probability
distribution will be
\begin{equation}
P'(b')= \sum_{b_1 b_2 y_1 y_2} \chi(b', b_1 b_2 y_1 y_2) P (b_1
b_2 | y_1 y_2)
\end{equation}
The only constraints on $\chi(b', b_1 b_2 y_1 y_2)$ are that it
satisfies the universality  and non-triviality  conditions
introduced in the previous section. We first consider
universality, which requires that $P'(b')$ is a valid probability
distribution for all initial two-box binary-input/binary-output
no-signalling states $P(b_1 b_2 | y_1 y_2)$.  I.e.
\begin{eqnarray}
0 \leq P'(b') \leq 1 \label{bipartite_prob_eqn1} && \\
\sum_{b'} P'(b') = 1 \label{bipartite_prob_eqn2}
\end{eqnarray}
It is easy to see that these properties are preserved under convex
combination. It is therefore only necessary to ensure that
(\ref{bipartite_prob_eqn1}) and (\ref{bipartite_prob_eqn2}) hold
for the 24 extremal states introduced in section \ref{box_sec}.
All other states can be represented as convex combinations of the
extremal states, and will therefore satisfy
(\ref{bipartite_prob_eqn1}) and (\ref{bipartite_prob_eqn2})
automatically.

Note that once $P'(0)$ is determined for each extremal state,
$P'(1)$ is fixed by equation (\ref{bipartite_prob_eqn2}) to be
$P'(1)=1-P'(0)$. Given any $\chi(0, b_1 b_2 y_1 y_2)$ satisfying
(\ref{bipartite_prob_eqn1}) it is always possible to find a
$\chi(1, b_1 b_2 y_1 y_2)$ satisfying (\ref{bipartite_prob_eqn1})
and (\ref{bipartite_prob_eqn2}) by defining $\chi(1, b_1 b_2 y_1
y_2) = 1/4-\chi(0, b_1 b_2 y_1 y_2)$), as this gives
\begin{eqnarray}
P'(1) &=& \sum_{b_1 b_2 y_1 y_2} \left( \frac{1}{4}-\chi(0, b_1
b_2 y_1 y_2) \right) P (b_1 b_2 | y_1 y_2) \nonumber \\ &=&
1-P'(0).
\end{eqnarray}
As all other choices of $\chi(1, b_1 b_2 y_1 y_2)$ must
give the same values for $P'(1)$, they are all equivalent. Hence
a coupler is completely specified by the 16 parameters $\chi(0,
b_1 b_2 y_1 y_2)$.

The class $\mathcal{X}$ of $\chi(0, b_1 b_2 y_1 y_2)$
distributions satisfying the universality constraint are those
which are consistent with the 48 linear inequalities that result
from applying equation (\ref{bipartite_prob_eqn1}) to the 16
extremal local states $P^L_{\alpha \beta \gamma \delta} (b_1 b_2 |
y_1 y_2 )$ and 8 extremal non-local states $P^N_{\alpha \beta
\gamma} (b_1 b_2 | y_1 y_2 )$. This approach yields a convex
polytope for $\mathcal{X}$ \footnote{We obtained the polytope
$\mathcal{X}$ using the program LRS, written by D.Avis.}. It has 9
dimensions and 82 vertices, as well as 7 linearities which have no
effect on the final probability distributions (and therefore
define an equivalence class of $\chi(0, b_1 b_2 y_1 y_2)$ which
would correspond to the same coupler).

Each point in the polytope $\mathcal{X}$ corresponds to a
different potential coupler, with the only remaining constraint on
them being that of non-triviality. The 82 extremal points of
$\mathcal{X}$  can all be expressed by $\chi(b, b_1 b_2 y_1 y_2)
\in \{0,1\}$, and can therefore be characterised by the values of
$\{b, b_1 b_2 y_1 y_2\}$ for which $\chi(b, b_1 b_2 y_1 y_2)=1$
(with all other coefficients being zero). Table
\ref{coupler_table} gives a representation of each extremal
$\chi(b, b_1 b_2 y_1 y_2)$ distribution in this way. For
simplicity, these extremal points have been divided into 5
classes, each of which are paramaterised by a subset of the binary
coefficients $\{\alpha, \beta, \gamma, \delta, \epsilon\} \in
\{0,1\}$.

\begin{table}  \begin{tabular}{|c|c|l|}
\hline Potential coupler  & Number
& Entries with $\chi(b', b_1 b_2 y_1 y_2)=1 $ \\
classes  & in class & ($\chi(b', b_1 b_2 y_1 y_2)=0$ otherwise)  \\
\hline
Deterministic $(\chi^D_{\alpha})$ & 2 & $\begin{array}{l} y_1=y_2=0, \\
 b'=\alpha \end{array}$
\\ \hline
One-sided $(\chi^{O}_{\alpha\beta\gamma})$ & 8 & $
\begin{array}{l} y_0=y_1=\alpha, \\ b'=b_{\beta} \oplus \gamma
\end{array}$
\\ \hline
XOR-gated $(\chi^X_{\alpha\beta\gamma})$ & 8 & $ \begin{array}{l}
y_1=\alpha, y_2=\beta, \\ b'=b_1 \oplus b_2 \oplus \gamma
\end{array}$
\\ \hline
AND-gated $(\chi^A_{\alpha\beta\gamma\delta\epsilon})$& 32 &
$\begin{array}{l} y_1=\alpha, y_2=\beta, \\ b'= (b_1\oplus
\gamma)(b_2\oplus \delta) \oplus \epsilon \end{array}$
\\ \hline
Sequential $(\chi^{S}_{\alpha\beta\gamma\delta\epsilon})$ & 32 & $
\begin{array}{l} y_{\alpha}=\beta, y_{(1 \oplus \alpha)} = b_{\alpha}
\oplus \gamma,  \\ b'= b_{(1 \oplus \alpha)} \oplus \delta
b_{\alpha} \oplus \epsilon \end{array}$
\\ \hline
\end{tabular} \caption{Table giving $\chi(b', b_1 b_2 y_1 y_2)$
for a representation of all 82 extremal points of $\mathcal{X}$,
corresponding to potential couplers (elements satisfying the
listed constraints are one, while all other elements are zero).
The potential couplers are divided into 5 classes, with the
members in each class parameterized by $\alpha, \beta, \gamma,
\delta, \epsilon \in \{0,1\}$. The second column shows the number
of different potential couplers in each class.
\label{coupler_table}}
\end{table}

\begin{figure}
 \epsfxsize=3.5truein
\centerline{\epsffile{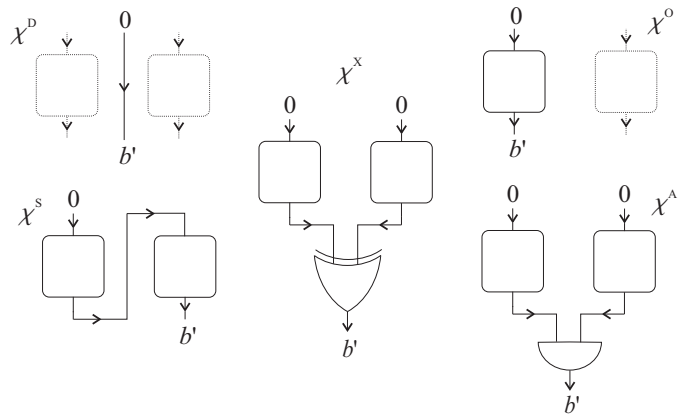}} \caption{Diagram showing how to
generate $\chi(b, b_1 b_2 y_1 y_2)$ for the standard potential
couplers ($\alpha=\beta=\gamma=\delta=\epsilon=0$) for each of the
5 classes defined in table \ref{coupler_table} using wiring and
logic gates (AND and XOR). In each circuit, the box on the left is
the box with input $y_1$. \label{wiringall_fig} }
\end{figure}

Perhaps surprisingly, all 82 extremal points of $\mathcal{X}$
\emph{fail} to satisfy the non-triviality constraint (iv). Each of
these potential couplers can be generated by applying wiring and
logic gates to Bob's two boxes (indeed, they represent every
inequivalent wiring of this type). A circuit diagram for the
standard potential coupler
($\alpha=\beta=\gamma=\delta=\epsilon=0$) in each class is shown
in figure \ref{wiringall_fig}. The remaining potential couplers in
each class can be obtained by adding NOT-gates to some or all of
the wires in the standard circuit, and/or by swapping the
positions of Bob's two boxes.
As every potential coupler in $\mathcal{X}$ can be realised by a
probabilistic mixture of these extremal potential couplers (and
hence a probabilistic wiring strategy for Bob), all of them fail
to satisfy the non-triviality constraint. There are therefore no
binary-output couplers for two binary-input/binary-output boxes.

Although we have so far only considered couplers with a binary
output $b'$, this result will hold for couplers with any range of
outputs $b\in \mathbb{N}$. If for some output $b=b_0$ a coupler
characterised by $\chi(b, b_1 b_2 y_1 y_2)$ existed, we could
always construct a binary-output coupler characterised by
$\chi'(b', b_1 b_2 y_1 y_2)$ which output $b'=0$ if $b=b_0$ and
$b'=1$ when $b\neq b_0$, by taking
\begin{eqnarray}
\chi'(b'=0, b_1 b_2 y_1 y_2) &=& \chi(b=b_0, b_1 b_2 y_1 y_2) \\
\chi'(b'=1, b_1 b_2 y_1 y_2) &=& \sum_{b \neq b_0} \chi(b, b_1 b_2
y_1 y_2)
\end{eqnarray}
As we have shown that there are no binary-output couplers, there
must also be no couplers with a larger output range. We can
therefore conclude that, in the case of two
binary-input/binary-output boxes, \emph{no couplers exist}. As
discussed in section \ref{non-locality1}, this means that it is
impossible to implement an analogue of entanglement swapping for
PR-states.

\section{Conclusions}

Considering correlation experiments in terms of abstract
black-boxes allows us to separate the information-theoretic
content of non-locality from the underlying physical detail.

For correlations between a  set of time-independent measurements,
no-signalling box-states represent every possibility which is
consistent with relativity. However, quantum theory also allows
the possibility of entanglement-swapping, in which non-local
correlations can be introduced between two subsystems which are
initially uncorrelated, by performing a joint measurement on two
subsystems with which they are entangled and announcing the
result.

We have shown that it is impossible to achieve an analogue of
entanglement-swapping between two bipartite non-local box-states
using a sequence of individual (yet conditional) box inputs. This
led us to introduce the concept of couplers, which are an analogue
of measurements with entangled eigenstates in quantum theory.

Under very general assumptions of universality (a coupler can be
applied to any box with the appropriate inputs and outputs),
completeness (a coupler acts identically on states with the same
probability distribution), and no-signalling, we found that any
allowed coupler must be linear. We then proceeded to investigate
the allowed couplers which act on two binary-input/binary-output
boxes. Perhaps surprisingly, we found that no couplers of this
type exist. In particular, this means that when Alice and Bob
share a PR-state, and Bob and Charlie share a PR-state, there is
no way that Bob can generate any non-local correlations between
Alice and Charlie.

As we have so far only explicitly considered couplers acting on
binary-input/binary-output boxes, it would be interesting to see
if couplers exist in more general cases. Of particular interest
would be the class of couplers which act on two
ternary-input/binary-output boxes and generate a two-bit output.
As qubit states can be characterised by measurements of the three
Pauli matrices, and Bell measurements have four outputs, this
would enable a closer analogy between the quantum case and that of
general correlated box states.

Couplers also correspond to a first step into the dynamics of
correlated boxes, transforming $n$ boxes in the initial state into
one effective box (with output $b'$ and no input) in the final
state. It is straightforward to generalise the constraints
introduced for couplers to apply to more general dynamical
processes (taking $n$ boxes in the initial state to $m$ boxes in
the final state). This opens the possibility for deeper studies of
the dynamics of correlated boxes (as desired in
\cite{no-bit-com}).

The results obtained so far for couplers suggest that by allowing
stronger non-local correlations than are attainable in quantum
theory, the dynamics of the model actually become weaker (to the
extent that no couplers exist for two binary-input/binary-output
boxes). This offers a possible insight into why such super-strong
correlations are not attainable in nature: In order to allow
richer dynamics.

\begin{acknowledgments} The authors would like to thank Serge
Massar, Lluis Massanes, Tobias Osbourne and Jonathan Barrett for
interesting discussions, and acknowledge support from the U.K.
Engineering and Physical Sciences Research Council (IRC ``Quantum
Information Processing'') and from the E.U. under European
Commission project RESQ (contract IST-2001-37559).
\end{acknowledgments}

\end{document}